\documentclass[11pt]{article}
\usepackage{exscale}\usepackage{graphicx}
\def\d{\mbox{\rm d}}
\begin{document}
\title{Using an old method of Jacobi to derive Lagrangians:
 a nonlinear dynamical system with variable coefficients}
 \author{M.C. Nucci  $\;$ and $\;$ K.M. Tamizhmani\footnote{Permanent address: Department of
Mathematics, Pondicherry University, Kalapet, Puducherry, 605 014,
INDIA, e-mail: tamizh@yahoo.com}}
\date{Dipartimento di Matematica
e Informatica,Universit\`a di Perugia, 06123 Perugia, Italy,
e-mail: nucci@unipg.it}

 \maketitle
 \begin{abstract}
We present a method devised by Jacobi to derive Lagrangians of any
second-order differential equation: it consists in finding a
Jacobi Last Multiplier. We illustrate the easiness and the power
of Jacobi's method by applying it to the same equation studied by
Musielak et al. with their own method [Musielak ZE, Roy D and
Swift LD.  Method to derive Lagrangian and Hamiltonian for a
nonlinear dynamical system with variable coefficients. Chaos,
Solitons \& Fractals, 2008;58:894-902]. While they were able to
find one particular Lagrangian after lengthy calculations, Jacobi
Last Multiplier method yields two different Lagrangians (and many
others), of which one is that found by Musielak et al, and the
other(s) is(are) quite new.
 \end{abstract}
\section{Introduction}
It should be well-known that the knowledge of a Jacobi Last
Multiplier always yields a Lagrangian of any second-order ordinary
differential equation \cite{JacobiVD}, \cite{Whittaker}. Yet many
distinguished scientists seem to be unaware of this classical
result. In this paper we present again the method of the Jacobi
Last Multiplier in order to compare the easiness and the power of
Jacobi's method with that proposed by Museliak et al
\cite{musetal08} for the same purpose. We have already presented
Jacobi's method in \cite{jlm05}. The references in \cite{jlm05}
and the papers \cite{ennity}-\cite{CP07jlmmech} may give an idea
of the many fields of applications yielded by Jacobi Last
Multiplier.

In \cite{musetal08} the authors searched for a Lagrangian of the
following second-order ordinary differential equation
\begin {equation}
\ddot x+b(x)\dot x^2+c(x)x=0 \label{eq1}
\end{equation}
with $b(x),c(x)$  arbitrary functions of the dependent variable
$x=x(t)$. After some lengthy calculations they found one
Lagrangian. In the present paper we apply Jacobi's method to
equation (\ref{eq1}) and show that many (an infinite number of)
Lagrangians can be easily derived.

This paper is organized in the following way. In  section 2, we
illustrate the Jacobi last multiplier and its properties
\cite{Jacobi 42}-\cite{JacobiVD}, its connection to Lie symmetries
\cite{Lie1874}, \cite{Lie 12 a}, and its link to the Lagrangian of
any second-order differential equations \cite{JacobiVD},
\cite{Whittaker}. Moreover we exemplify Jacobi's method by
determining the Lagrangians of two equations studied by Euler
\cite{Euler} and Jacobi himself \cite{Jacobi 45} for the purpose
of finding their multipliers. In section 3, we apply Jacobi's
method to equation (\ref{eq1}) and determine some of its many
Lagrangians. In section 4, we conclude with some final remarks.

 Here we employ ad hoc interactive programs
\cite{man2} written
in REDUCE language to calculate the Lie symmetry algebra of the equations we study.\\

\section{The method by Jacobi}
The method of the Jacobi last multiplier \cite{Jacobi 44
a}-\cite{JacobiVD}) provides a means to determine all the
solutions of the partial differential equation
\begin {equation}
\mathcal{A}f = \sum_{i = 1} ^n a_i(x_1,\dots,x_n)\frac {\partial
f} {\partial x_i} = 0 \label {2.1}
\end {equation}
or its equivalent associated Lagrange's system
\begin {equation}
\frac {\d x_1} {a_1} = \frac {\d x_2} {a_2} = \ldots = \frac {\d
x_n} {a_n}.\label {2.2}
\end {equation}
In fact, if one knows the Jacobi last multiplier and all but one
of the solutions, then the last solution can be obtained by a
quadrature. The Jacobi last multiplier $M$ is given by
\begin {equation}
\frac {\partial (f,\omega_1,\omega_2,\ldots,\omega_{n- 1})}
{\partial (x_1,x_2,\ldots,x_n)}
 = M\mathcal{A}f, \label {2.3}
\end {equation}
where
\begin {equation}
\frac {\partial (f,\omega_1,\omega_2,\ldots,\omega_{n- 1})}
{\partial (x_1,x_2,\ldots,x_n)} = \mbox {\rm det}\left [
\begin {array} {ccc}
\displaystyle {\frac {\partial f} {\partial x_1}} &\cdots &\displaystyle {\frac {\partial f} {\partial x_n}}\\
\displaystyle {\frac {\partial\omega_1} {\partial x_1}} & &\displaystyle {\frac {\partial\omega_1} {\partial x_n}}\\
\vdots & &\vdots\\
\displaystyle {\frac {\partial\omega_{n- 1}} {\partial x_1}}
&\cdots &\displaystyle {\frac {\partial\omega_{n- 1}} {\partial
x_n}}
\end {array}\right] = 0 \label {2.4}
\end {equation}
and $\omega_1,\ldots,\omega_{n- 1} $ are $n- 1 $  solutions of
(\ref {2.1}) or, equivalently, first integrals of (\ref {2.2})
independent of each other. This means that  $M$ is a function of
the variables $(x_1,\ldots,x_n)$ and  depends on the chosen $n-1$
solutions, in the sense that it varies as they vary. The essential
properties of the Jacobi last multiplier are:
\begin{description}
\item{ (a)} If one selects a different set of $n-1$ independent
solutions $\eta_1,\ldots,\eta_{n-1}$ of equation (\ref {2.1}),
then the corresponding last multiplier $N$ is linked to $M$ by the
relationship:
$$
N=M\frac{\partial(\eta_1,\ldots,\eta_{n-1})}{\partial(\omega_1,
\ldots,\omega_{n-1})}.
$$
\item{ (b)} Given a non-singular transformation of variables
$$
\tau:\quad(x_1,x_2,\ldots,x_n)\longrightarrow(x'_1,x'_2,\ldots,x'_n),
$$
\noindent then the last multiplier $M'$ of  $\mathcal{A'}F=0$ is
given by:
$$
M'=M\frac{\partial(x_1,x_2,\ldots,x_n)}{\partial(x'_1,x'_2,\ldots,x'_n)},
$$
where $M$ obviously comes from the $n-1$ solutions of
$\mathcal{A}F=0$ which correspond to those chosen for
$\mathcal{A'}F=0$ through the inverse transformation $\tau^{-1}$.
\item{ (c) } One can prove that each multiplier $M$ is a solution
of the following
 linear partial differential equation: \begin {equation}
\sum_{i = 1} ^n \frac {\partial (Ma_i)} {\partial x_i} = 0; \label
{2.5} \end {equation} \noindent viceversa every solution $M$ of
this equation is a Jacobi last multiplier.
\item{ (d) } If one
knows two Jacobi last multipliers $M_1$ and $M_2$ of equation
(\ref {2.1}), then their ratio is a solution $\omega$ of (\ref
{2.1}), or, equivalently,  a first integral of (\ref {2.2}).
Naturally the ratio may be quite trivial, namely a constant.
Viceversa the product of a multiplier $M_1$ times any solution
$\omega$ yields another last multiplier
$M_2=M_1\omega$.\end{description}
  Since the existence of a solution/first integral
is consequent upon the existence of symmetry, an alternate
formulation in terms of symmetries was provided by Lie \cite {Lie
12 a}. A clear treatment of the formulation in terms of
solutions/first integrals  and symmetries
 is given by Bianchi \cite {Bianchi 18 a}. If we know
$n- 1 $ symmetries of (\ref {2.1})/(\ref {2.2}), say
\begin {equation}
\Gamma_i =
\sum_{j=1}^{n}\xi_{ij}(x_1,\dots,x_n)\partial_{x_j},\quad i = 1,n-
1, \label {2.6}
\end {equation}
Jacobi's last multiplier is given by $M =\Delta ^ {- 1} $,
provided that $\Delta\not = 0 $, where
\begin {equation}
\Delta = \mbox {\rm det}\left [
\begin {array} {ccc}
a_1 &\cdots & a_n\\
\xi_{1,1} & &\xi_{1,n}\\
\vdots & &\vdots\\
\xi_{n- 1,1}&\cdots &\xi_{n- 1,n}
\end {array}\right]. \label {2.8}
\end {equation}
There is an obvious corollary to the results of Jacobi mentioned
above. In the case that there exists a constant multiplier, the
determinant is a first integral.  This result is potentially very
useful in the search for first integrals of systems of ordinary
differential equations.  In particular, if each component of the
vector field of the equation of motion is missing the variable
associated with that component, i.e., $\partial a_i/\partial x_i =
0 $, the last multiplier is a constant, and any other Jacobi Last
Multiplier is a first integral.

Another property of the Jacobi Last Multiplier is  its (almost
forgotten) relationship with the Lagrangian, $L=L(t,x,\dot x)$,
for any second-order equation
\begin{equation}
\ddot x=F(t,x,\dot x) \label{geno2}
\end{equation}
is \cite{JacobiVD}, \cite{Whittaker}
\begin{equation}
M=\frac{\partial^2 L}{\partial \dot x^2} \label{relMLo2}
\end{equation}
where $M=M(t,x,\dot x)$ satisfies the following equation
\begin{equation} \frac{{\rm d}}{{\rm d} t}(\log M)+\frac{\partial F}{\partial
\dot x} =0.\label{Meq}
\end{equation}
Then equation (\ref{geno2}) becomes the Euler-Lagrangian equation:
\begin{equation}
-\frac{{\rm d}}{{\rm d} t}\left(\frac{\partial L}{\partial \dot
x}\right)+\frac{\partial L}{\partial x}=0. \label{ELo2}
\end{equation}
The proof is given by taking the derivative of (\ref{ELo2}) by
$\dot x$ and showing that this yields (\ref{Meq}).
 If one knows a Jacobi last multiplier, then $L$ can be
easily obtained by a double integration, i.e.:
\begin{equation}
L=\int\left (\int M\, {\rm d} \dot x\right)\, {\rm d} \dot
x+f_1(t,x)\dot x+f_2(t,x), \label{lagrint}
\end{equation}
where $f_1$ and $f_2$ are functions of $t$ and $x$ which have to
satisfy a single partial differential equation related to
(\ref{geno2}) \cite{laggal}. As it was shown in \cite{laggal},
$f_1, f_2$ are related to the gauge function $g=g(t,x)$. In fact,
we may assume
\begin{eqnarray}
f_1&=&  \frac{\partial g}{\partial x}\nonumber\\
f_2&=& \frac{\partial g}{\partial t} +f_3(t,x) \label{gf1f2o2}
\end{eqnarray}
where $f_3$ has to satisfy the mentioned partial differential
equation and $g$ is obviously arbitrary.

In \cite{laggal} it was shown that if one knows several (at least
two) Lie symmetries of the second-order differential equation
(\ref{geno2}), i.e.
\begin {equation}
\Gamma_j =V_j(t,x)\partial_t+G_j(t,x)\partial_x, \quad j = 1,r,
 \label {gensym}
\end {equation}
 then many Jacobi Last Multipliers could be
derived by means of (\ref{2.8}), i.e.
\begin {equation}
{\displaystyle{\frac{1}{M_{nm}}}}=\Delta_{nm} = \mbox {\rm
det}\left [
\begin {array} {ccc}
1 &\dot x & F(t,x,\dot x)\\[0.2cm]
V_n &G_n &{\displaystyle{\frac{\d G_n}{\d t} -\dot x\frac{\d V_n}{\d t}}} \\[0.2cm]
V_m &G_m &{\displaystyle{\frac{\d G_m}{\d t} -\dot x\frac{\d V_m}{\d t}}}\\
\end {array}\right],\label {Mnm}
\end {equation}
with $(n,m=1,r)$, and therefore many Lagrangians  can be obtained
by means of (\ref{lagrint}).

In \cite{laggal} fourteen different Lagrangians\footnote{The
fourteen Lagrangians are independent from each other and not
related by any gauge function. They are derived from the
eight-dimensional Lie symmetry algebra which is admitted by any
linear second-order ordinary differential equations \cite{Lie 12
a}.} were derived even for an equation as controversial as the
damped linear harmonic oscillator, i.e.:
\begin{equation}
\ddot{x} + c\dot{x} + k x = 0, \label{1.3}
\end{equation}
which about 80 years ago was thought not to be derivable from a
variational principle by Bauer \cite{Bauer} and to be ``a
physically incomplete system'', namely in need of additional
equations\footnote{Bateman called them ``a complementary sets of
equations''.} by Bateman \cite{Bateman}.

\subsection{Two examples by Euler in \cite{Jacobi 45}}
In \cite{Jacobi 45}, Jacobi found his ``new multiplier'' for the
following class of second-order ordinary differential
equations\footnote{This is not Jacobi's original notation.}
studied by Euler \cite{Euler} [Sect. I,  Ch. VI, \S\S 915 ff.]:
\begin{equation}
\ddot x+\frac{1}{2}\frac{\partial \varphi}{\partial x}\dot
x^2+\frac{\partial \varphi}{\partial t}\dot x+B=0\label{Jeq}
\end{equation}
with $\varphi, B$ arbitrary functions of $t$ and $x$. Indeed
Jacobi derived that the multiplier of equation (\ref{Jeq}) is
given by:
\begin{equation}
M=e^{\varphi(t,x)}, \label{JM}
\end{equation}
as it is obvious from (\ref{Meq}). Then, Jacobi  presented two
examples of the class of equations (\ref{Jeq}), also studied by
Euler, to illustrate the use of his multiplier, namely that the
knowledge of one first integral and one multiplier yields
integration by quadrature. Here we report those equations for the
reader's convenience\footnote{Jacobi's original paper is in Latin,
but the mathematical formulas could be understood by any
mathematician.}. We also find the corresponding Lagrangian and
search for Lie symmetries of Euler's equation in order to describe
the link between Lie symmetries, Jacobi Last Multiplier, first
integrals, and of course Lagrangians. In fact it easy to derive a
Lagrangian of equation (\ref{Jeq}) by means by means of
(\ref{relMLo2}), i.e.
\begin{equation}
L=\frac{1}{2} e^{\varphi(t,x)}\dot x^2+f_1(t,x)\dot x+f_2(t,x)
\label{LagJ}
\end{equation}
with $f_1, f_2$ functions of $t$ and $x$ satisfying the following
equation:
\begin{equation}
\frac{\partial f_1}{\partial t} - \frac{\partial f_2}{\partial x}
=e^{\varphi(t,x)} B(t,x).
\end{equation}

\subsubsection{Example I}
The first example by Euler that Jacobi presented is the following
equation:
\begin{equation}
x^2 \ddot x+x\dot x^2+\beta x-\gamma t=0 \label{Euleq1}
\end{equation}
with $\beta$ and $\gamma$ arbitrary constants.   Jacobi obtained
the multiplier
\begin{equation}
M_1=x^2, \label{Eul1M1}
\end{equation}
by means of (\ref{JM}) and showed how to integrate equation
(\ref{Euleq1}).

We can use Jacobi's multiplier (\ref{Eul1M1}) to derive a
Lagrangian of equation (\ref{Euleq1}) by means of (\ref{relMLo2}),
i.e.
\begin{equation}
L_1=\frac{1}{2} x^2\dot x^2+f_1(t,x)\dot x+f_2(t,x) \label{Eul1L1}
\end{equation}
with $f_1, f_2$ functions of $t$ and $x$ satisfying the following
equation:
\begin{equation}
\frac{\partial f_1}{\partial t} - \frac{\partial f_2}{\partial x}
- \beta x + \gamma t=0. \label{Eul1f1f2}
\end{equation}
If we consider the transformation (\ref{gf1f2o2}) between
$f_1,f_2$ and the gauge function $g$, then equation
(\ref{Eul1f1f2}) becomes:
\begin{equation}
-\frac{\partial f_3}{\partial x} - \beta x + \gamma t=0,
\end{equation}
which can be easily integrated, i.e.:
\begin{equation}
f_3= \frac{1}{2}( - \beta x^2 + 2\gamma t x),
\end{equation}
Then $L_1$ can be written as follows:
\begin{equation}
L_1=\frac{1}{2} x^2\dot x^2+\frac{1}{2}( - \beta x^2 + 2\gamma t
x)+\frac{{\rm d}g}{{\rm d}t}. \label{Eul1L1f3}
\end{equation}
We remark that equation (\ref{Euleq1}) can be transformed into an
autonomous equation by considering the canonical variables $\tilde
t, \tilde x$ of the only Lie point symmetry admitted for any
$\gamma\neq 0$, i.e.
\begin{equation}
\Gamma=t\partial_t+x\partial_x,\quad \Longrightarrow \quad \tilde
t=\log(t), \; \tilde x=\frac{x}{t}.\label{Gamma1}
\end{equation}
Then equation (\ref{Euleq1}) becomes
\begin{equation}
\frac{{\rm d}\tilde x}{{\rm d}t} \frac{{\rm d}^2 \tilde x}{{\rm
d}t^2}= - \beta \tilde x + \gamma - \tilde x^3 - 3 \tilde
x^2\frac{{\rm d}\tilde x}{{\rm d}t} - \tilde x \left(\frac{{\rm
d}\tilde x}{{\rm d}t}\right)^2. \label{Euleq1t}
\end{equation}
with  Jacobi last multiplier and consequently Lagrangian
\begin{equation}
\tilde M_{1}=e^{\tilde t}\tilde x^2, \quad \Longrightarrow \quad
\tilde L_1=\frac{1}{2} e^{3\tilde t}\tilde x^2\left(\frac{{\rm
d}\tilde x}{{\rm d}t}\right)^2+\tilde f_1(\tilde t,\tilde
x)\frac{{\rm d}\tilde x}{{\rm d}t} +\tilde f_2(\tilde t,\tilde x)
\label{Eul1tL1} \end{equation}
 with $\tilde f_1, \tilde f_2$
satisfying the following equation:
\begin{equation}
\frac{\partial \tilde f_1}{\partial \tilde t} - \frac{\partial
\tilde f_2}{\partial \tilde x} - e^{3 \tilde t}\beta \tilde x +
e^{3 \tilde t}\gamma - e^{3 \tilde t}\tilde x^3=0.
\end{equation}
We find that the Lagrangian $L_1$ (\ref{Eul1L1}) does not admit
the  Lie point symmetry $\Gamma$  (\ref{Gamma1})
 as a Noether symmetry \cite{Noether}.

 If in equation (\ref{Euleq1}) we assume $\gamma=0$, i.e.:
 \begin{equation}
x^2 \ddot x+x\dot x^2+\beta x=0 \label{Euleq1lin}
\end{equation}
  then we find that it
 admits an eight-dimensional Lie symmetry algebra generated by the
following eight operators
\begin{eqnarray}
 \Gamma_1&=&\frac{1}{2 x} (\beta t^2+x^2)
  \left(2t x \partial_t+(x^2-\beta t^2)\partial_x\right)\nonumber\\
\Gamma_2&=& \frac{1}{ x}\left((3\beta t^2 x+x^3)\partial_t-2\beta^2 t^3\partial_x\right)\nonumber\\
\Gamma_3&=&\frac{t}{x}\left((2t x \partial_t+(x^2-\beta t^2)\partial_x\right)\nonumber\\
\Gamma_4&=&\frac{1}{ x} ( \beta t^2+x^2)\partial_x\nonumber\\
\Gamma_5&=&
\partial_t \label{Euleq1Gam} \\
\Gamma_6&=& \frac{t}{x}\left(x\partial_t-\beta t\partial_x\right)\nonumber\\
\Gamma_7&=&\frac{1}{ x}\,\partial_x\nonumber\\
\Gamma_8&=&\frac{t}{ x}\,\partial_x\nonumber
\end{eqnarray}
which means
 that equation (\ref{Euleq1lin}) is linearizable  by means of a point transformation
 \cite{Lie 12 a}. In order to find the
linearizing transformation we have to look for a two-dimensional
abelian intransitive subalgebra \cite{Lie 12 a}, and, following
Lie's classification of two-dimensional algebras in the real plane
\cite{Lie 12 a}, we have to transform it into the canonical form
(Type II) \begin{equation}\partial_{\tilde x},\;\;\;\;\;\tilde
t\partial_{\tilde x}\end{equation}  with $\tilde t$ and $\tilde x$
the new independent
 and dependent variables, respectively. We found that one such subalgebra
  is that generated by $\Gamma_7$
and $\Gamma_8$.
 Then, it is easy to derive that
 \begin{equation} \tilde t= t ,\;\;\;\;\;\tilde x= \frac{1}{2} x^2\end{equation}
and equation (\ref{Euleq1lin}) becomes
\begin{equation}
{{\rm d}^2 \tilde x \over {\rm d}\tilde t^2}=0.
\end{equation}
Now we can use the eight Lie point symmetries (\ref{Euleq1Gam}) to
generate fourteen different Jacobi last multipliers of equation
(\ref{Euleq1lin}) by means of (\ref{Mnm}), and therefore fourteen
different Lagrangians of equation (\ref{Euleq1lin}) by means of
(\ref{lagrint}) as it was shown in \cite{laggal}. Here we report
just three.

The Jacobi last multiplier $J_{35}$ which is derived from
$\Gamma_3$ and $\Gamma_5$, i.e.
\begin {equation}
J_{35}=-\frac{x^2}{(x\dot x+\beta t)(2 t x\dot x+\beta t^2-x^2)},
\end{equation}
yields the Lagrangian
\begin{eqnarray}
L_{35}&=& \frac{1}{2t(\beta t^2+x^2)}(-\beta t^2-2t x \dot
x+x^2)\log(\beta t^2+2tx \dot x-x^2) -\frac{1}{2t}\nonumber\\
&& +\frac{1}{\beta t^2+x^2}\log(\beta t+x \dot x)(\beta t+x \dot
x)+f_1\dot x+f_2,\nonumber\\
&& {\rm with} \quad \frac{\partial f_1}{\partial t} -
\frac{\partial f_2}{\partial x}=\frac{x}{t(\beta t^2+x^2)}.
\end{eqnarray}
The Jacobi last multiplier $J_{58}$ which is derived from
$\Gamma_5$ and $\Gamma_8$, i.e.
\begin {equation}
J_{58}=-\frac{x^2}{\beta t+ x\dot x},
\end{equation}
yields the Lagrangian
\begin{eqnarray}
L_{58}&=& -(\beta t+x\dot x)\log(\beta t+x \dot x)+\beta t+x \dot x+f_1\dot x+f_2,\nonumber\\
&&{\rm with} \quad \frac{\partial f_1}{\partial t} -
\frac{\partial f_2}{\partial x}=0.
\end{eqnarray}
The Jacobi last multiplier $J_{78}$ which is derived from
$\Gamma_7$ and $\Gamma_8$, i.e.
\begin {equation}
J_{78}={x^2}
\end{equation}
yields the Lagrangian
\begin{equation}
L_{78}= \frac{1}{2}x^2\dot x^2+f_1\dot x+f_2,\quad {\rm with}
\quad \frac{\partial f_1}{\partial t} - \frac{\partial
f_2}{\partial x}=\beta x.
\end{equation}
We note that $J_{78}$ corresponds to the multiplier (\ref{Eul1M1})
found by Jacobi.\\
We remind the reader that any ratio of two multipliers gives a
first integral of equation (\ref{Euleq1lin}), e.g.
\begin{eqnarray}
I_1&=&\frac{J_{58}}{J_{35}}=-x^2+2t x\dot x +\beta  t^2,\\
I_2&=&\frac{J_{78}}{J_{58}}=-x \dot x-\beta t,\\
I_3&=&\frac{J_{78}}{J_{35}}=(x \dot x+\beta t)(x^2-2t x\dot x
-\beta t^2).
\end{eqnarray}

\subsubsection{Example II}
The second example by Euler that Jacobi presented is the following
equation:
\begin{equation}
2x^3 \ddot x+x^2\dot x^2-\alpha x^2+\beta t^2-\gamma=0
\label{Euleq2}
\end{equation}
with $\alpha$, $\beta$ and $\gamma$ arbitrary constants.   Jacobi
obtained the multiplier
\begin{equation}
M_1=x, \label{Eul2M1}
\end{equation}
by means of (\ref{JM}) and showed how to integrate equation
(\ref{Euleq2}).

We can use Jacobi's multiplier (\ref{Eul2M1}) to derive a
Lagrangian of equation (\ref{Euleq2}) by means of (\ref{relMLo2}),
i.e.
\begin{equation}
L_2=\frac{1}{2} x\dot x^2+f_1(t,x)\dot x+f_2(t,x) \label{Eul2L2}
\end{equation}
with $f_1, f_2$ functions of $t$ and $x$ satisfying the following
equation:
\begin{equation}
2x^2\left(\frac{\partial f_1}{\partial t} - \frac{\partial
f_2}{\partial x} \right) + \alpha x^2 - \beta t^2 + \gamma=0.
\end{equation}
If we consider the transformation (\ref{gf1f2o2}) between
$f_1,f_2$ and the gauge function $g$, then:
\begin{equation}
f_3= \frac{1}{2x^2}(\alpha x^2 + \beta t^2 - \gamma),
\end{equation}
and $L_2$ can be written as follows:
\begin{equation}
L_2=\frac{1}{2} x\dot x^2+\frac{1}{2x^2}(\alpha x^2 + \beta t^2 -
\gamma)+\frac{{\rm d}g}{{\rm d}t}. \label{Eul2L2f3}
\end{equation}
We note that equation (\ref{Euleq2}) does not possess any Lie
point symmetry.

If we assume $\beta=\gamma=0$, then equation (\ref{Euleq2}), i.e.
\begin{equation}
2x^3 \ddot x+x^2\dot x^2-\alpha x^2=0 \label{Euleq2m}
\end{equation} admits a two-dimensional Lie symmetry algebra
generated by the following two operators
\begin{equation}
\Omega_1=t\partial_t+x\partial_x,\quad \quad
\Omega_2=\partial_t\,.
\end{equation}
 Then we can use these two symmetries  to generate another Jacobi
last multiplier of equation (\ref{Euleq2m}) by means of
(\ref{Mnm}), and therefore a different Lagrangian  by means of
(\ref{lagrint}). The Jacobi last multiplier $J_{12}$ which is
derived from $\Omega_1$ and $\Omega_2$, i.e.
\begin {equation}
J_{12}=\frac{2}{-\alpha+\dot x^2}, \label{J12}
\end{equation}
yields the Lagrangian
\begin{eqnarray}
L_{12}&=& -2\frac{\dot x}{\sqrt{\alpha}}\;{\rm
arctanh}\left(\frac{\dot x}{\sqrt{\alpha}}\right)
-\log\left(1-\frac{\dot x^2}{\alpha}\right)+f_1\dot x+f_2,\nonumber\\
&& {\rm with} \quad \frac{\partial f_1}{\partial t} -
\frac{\partial f_2}{\partial x}=-\frac{1}{x}.
\end{eqnarray}
Finally the ratio of the two multipliers (\ref{Eul2M1}) and
(\ref{J12}) is a first integral of equation (\ref{Euleq2m}), i.e.
\begin{equation}
I=\frac{M_1}{J_{12}}= \frac{1}{2} x(-\alpha+\dot x^2).
\end{equation}
If we assume $\alpha=\beta=\gamma=0$, then equation
(\ref{Euleq2}), i.e.
\begin{equation}
2x^3 \ddot x+x^2\dot x^2=0 \label{Euleq2mm}
\end{equation} admits an eight-dimensional Lie symmetry algebra,
and fourteen different Lagrangians can be derived.

\section{Application of the method by Jacobi to equation (\ref{eq1})}
It is obvious that equation (\ref{eq1}), for any $b(x)$ and
$c(x)$, is a particular case of the class of equations (\ref{Jeq})
studied by Jacobi. Therefore
 a Jacobi Last Multiplier (\ref{JM}) is already known, i.e.
\begin{equation}
M_1=e^{2 P_b(x)}, \quad \quad {\rm with} \quad P_b(x)=\int b(x)\,
dx, \label{M1mus}
\end{equation}
and the corresponding Lagrangian (\ref{LagJ}) is
\begin{equation}
L_1=\frac{1}{2} e^{2 P_b(x)}\dot x^2+f_1(t,x)\dot x+f_2(t,x)
 \label{L1mus}
\end{equation}
with $f_1, f_2$ functions of $t$ and $x$ satisfying the following
equation:
\begin{equation}
\frac{\partial f_1}{\partial t} - \frac{\partial f_2}{\partial x}
=e^{2 P_b(x)} c(x) x. \label{ff1mus}
\end{equation}
The Lagrangian derived by Museliak et al. \cite{musetal08}, after
lengthy calculations, i.e.
\begin{equation}
L=\frac{1}{2} e^{2 P_b(x)}\dot x^2-\int e^{2 P_b(x)} c(x) x\, dx,
\end{equation}
 is  a subcase of the Lagrangian (\ref{L1mus}),
  with $f_1=0$ and $f_2=f_2(x)=\int e^{2 P_b(x)} c(x) x\, dx$, which is
  an  obvious particular solution of (\ref{ff1mus}).
Actually we can derive another Lagrangian of (\ref{eq1}). In fact
we note that (\ref{eq1}) admits one trivial Lie point symmetry for
any $b(x)$ and $c(x)$, i.e. $\Gamma=\partial_t$, which is also a
Noether's symmetry \cite{Noether} for the Lagrangian
(\ref{L1mus}). Therefore a first integral can be easily
obtained\footnote{This energy-type first integral was obtained in
\cite{musetal08}.} from Noether's theorem \cite{Noether}, i.e.:
\begin{equation}
I_1=\frac{1}{2} e^{2 P_b}\dot x^2 + \int  e^{2 P_b} c(x)x\, dx.
\label{Imus}
\end{equation}
Let us use the  property (d) of the Jacobi last multiplier. If one
knows a Jacobi last multiplier $M_1$ in (\ref{M1mus}) and a first
integral  $I_1$ in (\ref{Imus}) of equation (\ref{eq1}), then
their product is another Jacobi last multiplier, i.e.
\begin{equation}
M_2=M_1 I_1 = \frac{1}{2} \,e^{2 P_b}\left(e^{2 P_b}\dot x^2 +
\int e^{2 P_b} c(x)x\, dx\right) \label{M2mus}
\end{equation}
Consequently  we are able to obtain a second Lagrangian of
equation (\ref{eq1}) for any $b(x)$ and $c(x)$, i.e.
\begin{equation}
L_2=\frac{1}{24}\,e^{2 P_b}\dot x^2\left(e^{2 P_b}\dot x^2 + 12
\int  e^{2 P_b} c(x)x\, dx\right) +f_1(t,x)\dot x+f_2(t,x)
\label{L2mus}
\end{equation}
with $f_1, f_2$ functions of $t$ and $x$ satisfying the following
equation:
\begin{equation}
\frac{\partial f_1}{\partial t} - \frac{\partial f_2}{\partial x}
=e^{2 P_b(x)} c(x)\int  e^{2 P_b} c(x)x\, dx.
\end{equation}
This Lagrangian admits $\Gamma=\partial_t$ as a Noether's symmetry
and the corresponding first integral is just the square of $I_1$
in (\ref{Imus}).\\
We can keep using property (d) to derive more and more Jacobi last
multipliers and therefore Lagrangians of equation (\ref{eq1}). In
fact other Jacobi last multipliers can be obtained by simply
taking any function of the first integral $I_1$ in (\ref{Imus})
and multiplying it for either $M_1$ in (\ref{M1mus}) or $M_2$ in
(\ref{M2mus}), and so on ad libitum. For example, we may take
$I_1^2$, and then obtain another Jacobi last multiplier by taking
the product of $M_1$ in (\ref{M1mus}) and $I_1^2$, i.e.:
\begin{equation}
M_3=M_1 I_1^2 = \frac{1}{4} \,e^{2 P_b}\left(e^{2 P_b}\dot x^2 +
\int e^{2 P_b} c(x)x\, dx\right)^2
\end{equation}
which yields the following third Lagrangian of equation
(\ref{eq1})
\begin{eqnarray}
L_3&=&\frac{1}{120}\,e^{2 P_b}\dot x^2\left(e^{4 P_b}\dot x^4 +10
e^{2 P_b}\dot x^2  \int  e^{2 P_b} c(x)x\, dx+ 60 \left(\int  e^{2
P_b} c(x)x\, dx\right)^2\right)\nonumber\\ &&+f_1(t,x)\dot
x+f_2(t,x) \label{L3mus}
\end{eqnarray}
with $f_1, f_2$ functions of $t$ and $x$ satisfying the following
equation:
\begin{equation}
\frac{\partial f_1}{\partial t} - \frac{\partial f_2}{\partial x}
=e^{2 P_b(x)} c(x)x\left(\int  e^{2 P_b} c(x)x\, dx\right)^2.
\end{equation}

\section{Final remarks}
The purpose of the present paper is to exemplify once again the
method of Jacobi for finding Lagrangians, and to stress its strong
connection with Lie symmetries and Noether symmetries.  We
advocate Jacobi Last Multiplier as an essential tool for studying
nonlinear dynamical systems.
 As (apparently) stated by Henry S. Truman:\\
{\em There is nothing new in the world, except the history one
does not know}.

\section*{Acknowledgements}
K.M.T. thanks Professor M.C. Nucci and the Dipartimento di
Matematica e Informatica,
 Universit\`a di Perugia, for hospitality and the provision of facilities
  while this work was prepared, and gratefully acknowledges the support of the Italian Istituto
Nazionale Di Alta Matematica ``F. Severi'' (INDAM), Gruppo
Nazionale per la Fisica Matematica (GNFM), Programma Professori
Visitatori.


\begin{thebibliography}{99}
\bibitem{Bateman}  Bateman H. On dissipative systems and related
variational principles. Phys Rev, 1931;38:815-819.
\bibitem{Bauer} Bauer PS. Dissipative dynamical systems I.
 Proc Nat Acad Sci, 1931;17:311-314.
\bibitem {Bianchi 18 a}
 Bianchi L. Lezioni sulla teoria dei gruppi continui finiti
di trasformazioni. Pisa: Enrico Spoerri; 1918.
\bibitem{Euler}
 Euler L. E366 -  Institutionum calculi integralis, Volumen
Secundum. Petropli impensis academiae imperialis scientiarum;
1769.
\bibitem{Jacobi 42}
 Jacobi CGJ. Sur un noveau principe de la m\'ecanique
 analytique.  C R Acad Sci Paris, 1842;15:202--205.
 \bibitem {Jacobi 44 a}
Jacobi CGJ. Sul principio dell'ultimo moltiplicatore, e suo uso
come nuovo principio generale di meccanica.  Giornale Arcadico di
Scienze, Lettere ed Arti, 1844;99:129--146.
\bibitem {Jacobi 44 b}
Jacobi CGJ.  Theoria novi multiplicatoris systemati \ae quationum
differentialium vulgarium applicandi. J Reine Angew Math, 1844;
27:199--268.
\bibitem {Jacobi 45}
Jacobi CGJ. Theoria novi multiplicatoris systemati \ae quationum
differentialium vulgarium applicandi. J Reine Angew Math,
1845;29:213-279 and 333-376.
 \bibitem{JacobiVD}  Jacobi CGJ.
 Vorlesungen \"uber Dynamik. Nebst f\"unf hinterlassenen Abhandlungen
 desselben herausgegeben  von A.Clebsch. Berlin: Druck und Verlag von Georg
 Reimer; 1886.
 \bibitem{Lie1874}  Lie S. Veralgemeinerung und neue Verwerthung der Jacobischen
Multiplicator-Theorie. Fordhandlinger i Videnokabs - Selshabet i
Christiania, 1874; pp. 255-274.
 \bibitem {Lie 12 a}
 Lie S. Vorlesungen \"uber Differentialgleichungen mit
bekannten infinitesimalen Transformationen. Leipzig: Teubner;
1912.
  \bibitem{musetal08}
 Musielak ZE, Roy D and Swift LD.  Method to derive Lagrangian and
Hamiltonian for a nonlinear dynamical system with variable
coefficients. Chaos, Solitons \& Fractals, 2008;58:894-902.
\bibitem{Noether}
  Noether E. Invariante Variationsprobleme. Nachr d K\"onig Gesellsch d Wiss zu
 G\"ottingen Math-phys Klasse, 1918;235-257.
\bibitem{man2}
 Nucci MC. Interactive REDUCE programs for calculating Lie
point, non-classical, Lie-B\"{a}cklund, and approximate symmetries
of differential equations: manual and floppy disk, in CRC Handbook
of Lie Group Analysis of Differential Equations, Vol. 3: New
Trends in Theoretical Developments and Computational Methods, ed.
 Ibragimov NH. Boca Raton: CRC Press; 1996, pp. 415-481.
 \bibitem{ennity}
Nucci MC and Leach PGL.
 Jacobi's last multiplier and the complete symmetry group of the Euler-Poinsot
 system.
  J Nonlinear Math Phys,  2002;9-s2:110-121.
\bibitem{JLM}
Nucci MC and Leach PGL.  Jacobi's last multiplier and symmetries
for the Kepler problem plus a lineal story. J Phys A: Math Gen,
2004;37:7743-7753.
\bibitem{jlm05}
Nucci MC.  Jacobi last multiplier and Lie symmetries:
 a novel application of an old relationship.
  J Nonlinear Math Phys, 2005;12:284-304.
  \bibitem{jlmcsepeq}
Nucci MC and Leach PGL. Jacobi's last multiplier and the complete
symmetry group of the Ermakov-Pinney equation. J Nonlinear Math
Phys, 2005;12:305-320.
\bibitem{gallipoli04}
Nucci MC.  Let's Lie: a miraculous haul of fishes. Theor Math
Phys, 2005;144:1214-1222.
\bibitem{PVI}
Nucci MC and Leach PGL.  Fuch's solution of Painlev\'e VI equation
by means of Jacobi last multiplier. J Math Phys, 2007;48:013514 (7
pages)
 \bibitem{gallipoli06}
Nucci MC.  Jacobi last multiplier, Lie symmetries, and hidden
linearity: ``goldfishes" galore. Theor Math Phys,
2007;151:851--862.
\bibitem{laggal}
Nucci MC and Leach PGL.  Lagrangians galore. J Math Phys,
2007;48:123510 (16 pages).
\bibitem{PInceeq}
 Nucci MC.
 Lie symmetries of a Panlev\'e-type equation without Lie
 symmetries. J Nonlinear Math Phys, 2008;15: (to appear)
 \bibitem{jlmschqm}
Nucci MC and Leach PGL. Gauge variant symmetries for the
Schr\"odinger equation.  Il Nuovo Cimento B, 2008;123: (to appear)
\bibitem{CP07Rao1JMP}
Nucci MC and Leach PGL. Jacobi last multiplier and Lagrangians for
multidimensional linear systems. J Math Phys, 2008;49: (to appear)
\bibitem{CP07jlmmech}
Nucci MC and Leach PGL. The Jacobi Last Multiplier and
Applications in Mechanics.  Physica Scripta, 2008; (to appear)
\bibitem{Whittaker}
 Whittaker ET.  A Treatise on the Analytical Dynamics of
Particles and Rigid Bodies. Cambridge: Cambridge University Press:
1988 (First ed. 1908).


\end{thebibliography}
 \end{document}